\documentclass[twocolumn,showpacs,preprintnumbers,amsmath,amssymb]{revtex4}

\usepackage{graphicx}
\usepackage{dcolumn}
\usepackage{bm}

\begin{document}


\title{Electron spin quantum beats in positively charged quantum dots: nuclear field effects}

\author{L. Lombez, P.-F. Braun, X. Marie } 
 \author{P. Renucci, B. Urbaszek} 
\author{T. Amand}
 \email{amand@insa-toulouse.fr}
\affiliation{Laboratoire de Physique et Chimie des Nano-Objets, UMR 5215, INSA-CNRS-UPS, 135 Avenue de Rangueil, 31077 Toulouse CEDEX 4, France }%

\author{O. Krebs, P. Voisin}
\affiliation{Laboratoire de Photonique et Nanostructures, Route de Nozay, 91460 Marcoussis, France}%

\date{\today}

\begin{abstract}
We have studied the electron spin coherence in an ensemble of positively charged InAs/GaAs quantum dots. In a transverse magnetic field, we show that two main contributions must be taken into account to explain the damping of the circular polarization oscillations. The first one is due to the nuclear field fluctuations from dot to dot experienced by the electron spin. The second one is due to the dispersion of the transverse electron Land\'e g-factor, due to the inherent
inhomogeneity of the system, and leads to a field dependent contribution to the damping. We have developed a model taking into account both contributions, which is in good agreement with the experimental data. This enables us to extract the pure contribution to dephasing due to the nuclei. 
\end{abstract}

\pacs{Valid PACS appear here}
\maketitle

\section{Introduction}

A single carrier spin in a single quantum dot represents a potential candidate for q-bit implementation in a solid state environment, in view of applications in the fields of spintronics and quantum information processing \cite{DDA,DPDV,Imamoglu,Zutic}. Quantum dots (QDs) are indeed attractive with respect to the general criteria for quantum computers \cite{Criteres}, since long spin relaxation times \textit{$T_1$} have been measured in neutral QDs, in the millisecond range for longitudinal magnetic field between 4 and 8 T \cite{Kroutvar, Elzerman}. This is made possible due to the inhibition of the random interactions leading to spin relaxation and decoherence in bulk \cite{Optical} and quantum well semiconductor structures. However, one of the major difficulties to achieve quantum manipulations is decoherence due to interactions with an uncontrollable environment. Long spin decoherence times \textit{$T_2$} are indeed demanded in order to be able to achieve enough quantum manipulations during the spin lifetime $T_1$, and this requirement turns out to be the most stringent one. In principle, the dephasing rate due to spin-orbit coupling may reach values of the order of \textit{$T_1$} \cite{Golovach}. Recent experiments have demonstrated a trend in this direction, where \textit{$T_2$} of the order of 1 $\mu s$ could be measured at low temperature, allowing possible applications based on robust quantum coherence within an ensemble of dots \cite{Petta, Greilich-science}.

In quantum dots, two dephasing processes are still efficient, and even enhanced by confinement: the exchange interaction between charges, and the hyperfine interaction. The first one is responsible for various phenomena, such as \textit{e. g.} optically active exciton state splitting, neutral exciton quantum beats and the appearance of negative circular polarization for negatively charged excitons under non resonant excitation \cite{Steel,Tartakovskii,Senes,Laurent}. The second one, the hyperfine interaction of localized electrons with the QD nuclei, leads to very efficient spin dephasing \cite{Merkulov, Khaetskii, Semenov, Akimov}. This interaction effect has been observed previously in p-doped InAs/GaAs QDs for the ground state of the positively charged exciton (also called trion) \textit{$X^+$} by time resolved photoluminescence (TRPL)\cite{Braun}, and more recently in n-doped QDs for the resident electron using time resolved Kerr pump-probe spectroscopy \cite{Greilich, Greilich-science}. Three distinct time scales are relevant in describing the electron-nuclei spin system evolution \cite{Merkulov}: the first time corresponds to the electron spin precession around the frozen nuclear field fluctuations due to the QD nuclei (the typical dephasing time is of the order of 1 ns for GaAs QDs containing $10^5$ nuclei); the second one is controlled by nuclear spin precession in the hyperfine field of the localized electron (the typical time is of the order of 1 $\mu s$), and the last one is the nuclear spin relaxation due to dipole-dipole interaction with nuclei in the vicinity of the QDs (the typical time is of the order of 100 $\mu s$). During the first two stages, the coherence of the electron-nuclear spin system is preserved, while during the last one it is not, since the dipole-dipole interaction does not conserve the total nuclear spin. In ref.\cite{Petta} and \cite{Braun}, it is essentially the first dephasing stage which is observed, leading to an estimate of the electron spin ensemble dephasing time $T_2^*$ in the inhomogeneous nuclear field. The longer spin coherence times measured in ref. \cite{Petta,Greilich-science} are based respectively on spin echo technique or mode locking of electron spin coherences to suppress the hyperfine induced dephasing, and are presumably limited by the nuclear spin dephasing time. 

In this work, we investigate the first stage of electron spin dephasing in an ensemble of InAs/GaAs QDs in the presence of an in plane magnetic field. To achieve this aim, we study the electron spin coherence (ESC) in time resolved photoluminescence spectroscopy of positively charged excitons $X^+$. Exciting QDs containing a single doping hole with an optical pulse results in the formation of a trion $X^+$ which, in its ground state, consists of a hole spin singlet in the highest valence states and a single electron spin in the lowest conduction orbital state:
\begin{eqnarray}	\left|X^+,\pm\frac{1}{2}\right\rangle=\frac{1}{\sqrt{2}}\left(\left|\frac{3}{2},-\frac{3}{2}\right\rangle-\left|-\frac{3}{2},\frac{3}{2}\right\rangle\right)\otimes\left|\pm\frac{1}{2}\right\rangle
\end{eqnarray}
where $\left|\pm\frac{3}{2}\right\rangle$, $\left|\pm\frac{1}{2}\right\rangle$ represent respectively the projection of the heavy hole and conduction electron angular momentum on the quantification axis $Oz$, taken normal to the sample surface. 
Hence, the electron-hole exchange interaction which is efficient within neutral QD excitons \cite{Bayer,Senes} is cancelled out \cite{Braun, Eble}. The time and polarization resolved photoluminescence (TRPL) signal is thus a direct probe of the unpaired electron spin dynamics during the radiative time. Note that in this approach, when exciting the dots within the wetting layer (WL), the holes lose their spin polarization before they are captured by the QDs \cite{Laurent,Cortez} due to efficient spin relaxation in the WL \cite{Damen}. In addition, the spin coherence of localized holes that might be generated by the laser pulse, in a symmetrical way as for negatively charged dots \cite{Economou,Dutt,Greilich}, will be cancelled out after the formation of the hole singlet in the trion ground state. Finally, the hyperfine interaction of the resident hole spin with the nuclei is negligible, due to the p-symmetry of the periodic part of the Bloch function \cite{Abragam,Gryncharova}. Note also that, for n-doped QDs, the Kerr resonant pump-probe approach is very appropriate to study long term resident electron spin evolution, particularly after the radiative recombination \cite{Greilich-science,Greilich}; however, the response at short time delay is more complex to analyze using this technique: both the dots with a photogenerated trion and the dots with a single carrier bearing an optically generated spin coherence contribute to the Kerr probe signal, leading to a complex interference pattern under transverse magnetic field (due to non zero transverse g-factor of the hole in QDs \cite{Bayer}). The TRPL experiment is then well adapted to make an accurate description of the unpaired electron spin coherence of $X^+$ during the first step of its dephasing by nuclear spins.
As we observe the average electron spin $\left\langle S_z(t)\right\rangle$ in an ensemble of dots, the decay of the oscillation amplitude can be described by a characteristic spin dephasing time $T_2^*$; in a classical view, this is the decoherence time of the spin ensemble during the precession around the applied transverse magnetic field, taking into account the different inhomogeneities of the sample. 

We have developed a theoretical model to describe the quantum beats observed experimentally. We observe that two contributions participate to the spin dephasing: the first one is due to the nuclear field fluctuations. However, this contribution alone cannot explain the observed dephasing: the second contribution is due to static Land\'e g-factor fluctuations from dot to dot \cite{Dutt, Greilich}. Finally, variations of QD doping level have also to be taken into account. By fitting our theoretical model to the experimental data, we can extract the contribution to the dephasing due to the hyperfine interaction alone. The hyperfine interaction is also responsible for the dephasing in a single quantum dot, when averaging over a large number of nuclear field fluctuations. This is the case when the signal integration time is much larger than nuclear spin relaxation time due to the dipole-dipole interaction. 

\section{SAMPLES AND EXPERIMENTAL SET-UP}
The samples used for this work consist of 10 planes of self assembled InAs QDs grown by Stranski-
Krastanow method, and separated by 30 nm of GaAs. A beryllium delta doping layer
is located $15nm$ below each wetting layer. Several samples have been investigated with different nominal doping corresponding roughly to one hole per dot on average. We show here the results for two typical samples : sample 1 with a nominal doping of  $N_A=5\times10^{10}cm^{-2} $, and sample 2 with a nominal doping of $N_A=15\times10^{10}cm^{-2} $. The samples are mounted in a cryostat cooled at a
temperature of 15K. The experiment is performed in Voigt geometry, the external transverse magnetic field $\textbf{B} = B_x \textbf e_x$ being oriented along the $Ox$ axis in the QDs plane. The QDs are excited with 1.5ps pulses from a mode-
locked Ti-sapphire laser with a repetition rate of 80 MHz. Circularly polarized light ($\sigma +$) propagating along the growth axis $Oz$ generates a coherent superposition of $|X^+,+1/2>_x$ and $|X^+,-1/2>_x$ trion eigenstates. The excitation beam is focused on a spot size of $100\mu m$ diameter with an average power of $1mW$. We did not observe any change in the PL circular polarization dynamics for excitation powers up to 5mW. The PL intensities co-polarized ($I^+$) and counter-polarized ($I^-$) with the excitation laser are dispersed by a monochromator and then recorded using a S1 photocathode Hamamatsu Streak Camera with an overall time-resolution of 30 ps. We measure the circular polarization degree of the photoluminescence $P_c=(I^+ - I^-)/(I^+ + I^-)$  , which corresponds directly to the electron average spin component $\left\langle S_z(t)\right\rangle = -P_c/2$ along $Oz$ \cite{Optical}.

\section{Results and discussions}

\begin{figure}
\includegraphics[width=0.47\textwidth]{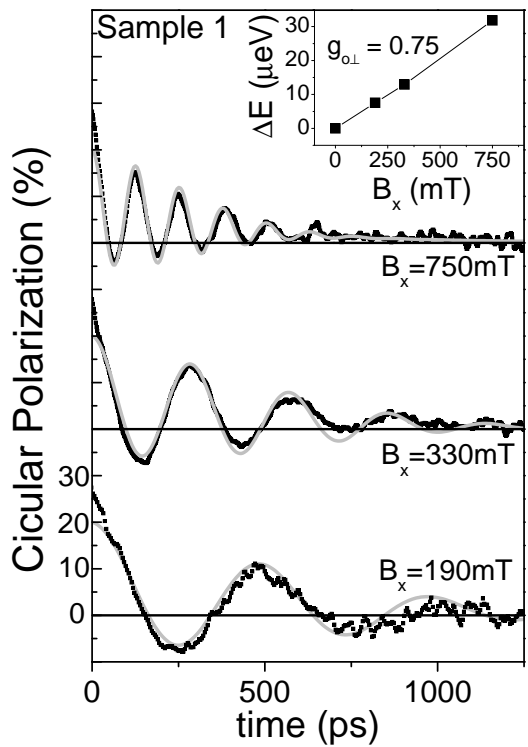}
\caption{\label{fig:5269} Sample 1 ($N_A=5\times10^{10}cm^{-2} $):  time-resolved circular polarization decay of the $X^+$ trion under $\sigma+$ excitation for three different magnetic field strengths $B_x$. Black dotted lines represent experimental curves, grey lines are theoretical curves (see text). Inset: Zeeman splitting as a function of $B_x$ .  }
\end{figure}

We present first the  circular polarization ($P_c$) dynamics when increasing the magnetic field $B_x$ up to 750mT. 
The excitation energy ($E_{ex} = 1.44eV$ ) corresponds to the lowest states of the wetting layer (WL)
\cite{Cortez}. Under these excitation conditions we assume, consistently with previous experiments, that the electron maintains its spin orientation during the capture in QDs \cite{Cortez}. For sample 1, figure \ref{fig:5269} presents the time evolution of the circular polarization $P_c(t)$ for three different magnetic fields $B_x$. At zero magnetic field, we find positive circular polarization (not shown) confirming that the QDs are p-doped \cite{Braun}. We recall that for neutral QDs the anisotropic exchange interaction yields linearly polarized neutral exciton eigenstates and under non resonant excitation, the circular polarization value is lower than 3\% \cite{Paillard, Eble}. For n-doped QDs, the eigenstates are circular, but the polarization rate is negative due to interplay of anisotropic exchange interaction and Pauli blocking \cite{Cortez, Laurent, Akimov, Oulton}. Moreover, as we can observe luminescence under strictly resonant excitation (not shown here), we deduce that there are on average less than 2 holes per dot. We assume first that all the dots contain a single resident doping hole (this point will be discussed later). After relaxation, the optically excited and the doping holes will finally form a singlet in the trion ground state, which recombines radiatively. 

\begin{figure}
\includegraphics[width=0.47\textwidth]{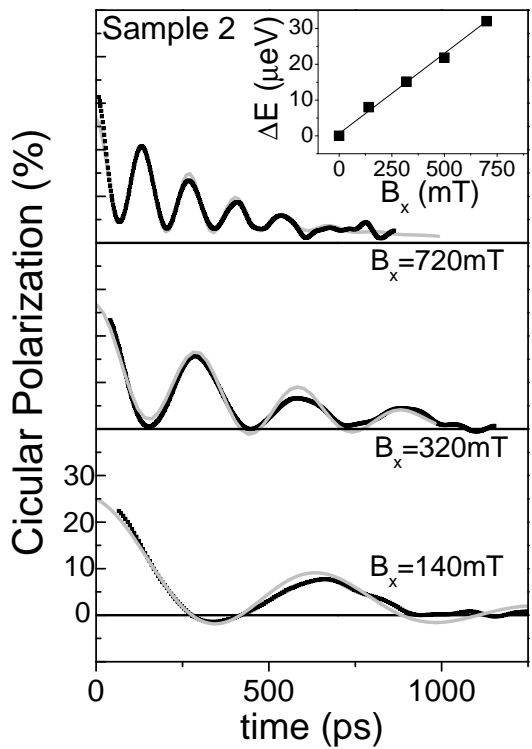}
\caption{\label{fig:5274} Sample 2 ($N_A=15\times10^{10}cm^{-2}$): same as Fig.\ref{fig:5269} but increasing the doping level of the structure.
}
\end{figure}

The oscillation period observed, corresponding to the $X^+$ ground state Zeeman splitting $\Delta E$, decreases when the magnetic field increases. 
The inset in Fig.\ref{fig:5269} shows the linear dependence: $\Delta E=g_{\bot0} \mu_B B_x$. 
We can thus find the average transverse electron Land\'e g-factor $\left|g_{\bot0}\right|\approx 0.75\pm 0.05$. Note that, as in quantum wells \cite{Ivchenko,Lejeune,Malinowski,Hendorfer}, the electron effective Land\'e g-factor is strongly anisotropic \cite{Bayer2}. 

From the curves $P_c(t)$ plotted in Fig.\ref{fig:5269} we also see that despite the relatively high magnetic field
value, the ESC decays with a typical time $T_2^*$ of about 300 ps. Increasing the magnetic field also leads to an increase of the damping of the polarization oscillations. Furthermore, an unusual asymmetry of spin quantum beats is observed, the negative extrema of the circular polarization being smaller in absolute value than the positive ones, in contrast with what is commonly observed \cite{Greilich,Heberle,Amand,Senes}. Comparing Fig.\ref{fig:5269} and Fig.\ref{fig:5274}, we see that this asymmetry increases with the doping level of the structure.

To explain these observations, we  discuss first the role of the
interaction of the trion with nuclei. Secondly, we will explain the dependence of the beats damping on $B_x$ magnitude and the reason for the asymmetry. Finally, we will show the agreement between the experimental
data and the different theoretical parameters used in this paper.

As the correlation time of the nuclear field is much longer than the $X^+$ trion lifetime \cite{Braun}, we start from the model of the "frozen" configuration of nuclear spins \cite{Merkulov}. In a first step, we only consider the electron spin dephasing originating from the random distribution of the nuclear hyperfine field in the dots. 

In order to determine the equation of motion of a spin in a dot, we adopt a smilar approch as in Ref \cite{Merkulov}. However here, care has to be taken in order to take into account the g-factor anisotropy. It is thus more convenient to use the precession vector picture rather than the effective magnetic field picture. The Hamiltonian of an electron in a dot is:
\begin{eqnarray}
	H=H_N+H_Z 
\end{eqnarray}\begin{eqnarray}	
	H=v_0 \sum_i A_i\left|\Psi(\textbf R_i)\right|^2 \hat{\textbf I}_i.\hat{\textbf S}+(~\mu_B\textbf{g}\textbf B_{ext}).\hat{\textbf{S}}
\end{eqnarray}
where the first term is the usual hyperfine contact Hamiltonian, and the second is the Zeeman hamiltonian. Here $\textbf{g}$ is the electron g-factor tensor, $\hat{\textbf I}_i$ and $\hat{\textbf S}$ are writen in $\hbar$ units, and $v_0$ is the volume of the unit cell. The hyperfine interaction constants $A_i$ are all of the order of $50\mu eV$ considering two atoms per elementary cell \cite{Urbaszek}. We can thus write:

\begin{eqnarray}
	H=\hbar~\hat{\bm\Omega}.\hat{ \textbf{S}}=\hbar~(\hat{\bm\Omega}_N + \hat{\bm\Omega}_{ext}).\hat{\textbf{S}}
\end{eqnarray}

where $\hat{\bm\Omega}_N=v_0/\hbar \sum_i A_i\left|\Psi(\textbf R_i)\right|^2 \hat{\textbf I}_i$ is the precession vector operator and $\hat{\bm\Omega}_{ext}=\mu_B\textbf{g} \textbf B_{ext}/\hbar$.
The quantum mechanical average of $\hat{\bm\Omega}_N$ on a given nuclear configuration is:
 
\begin{eqnarray}
\label{ON}
	\bm\Omega_N=\left\langle \hat{\bm\Omega}_N \right\rangle=\frac{v_0}{\hbar}\left\langle  \sum_i A_i\left|\Psi(\textbf R_i)\right|^2 \hat{\textbf I}_i\right\rangle
\end{eqnarray}

In the "frozen fluctuation" approach, $\bm\Omega_N$ evolves with time much slower than the electron average spin $\left\langle \hat{\bf{S}}(t)\right\rangle$. The electron sees a distribution of $\bm\Omega_N$, whose magnitude and direction are randomly distributed over the QD ensemble, described by an isotropic gaussian probability density distribution function :

\begin{eqnarray}
\label{OmegaN}
W(\bm\Omega_N)=\frac{1}{\pi^{3/2}\Delta_{\Omega_N}^3} exp\left(-\frac{\bm\Omega_N^2}{\Delta_{\Omega_N}^2}\right)
\end{eqnarray}

where $\Delta_{\Omega_N}$ represents the dispersion of the precession vector. The fluctuation of $\bm\Omega_N$ can be readily obtained assuming that the nuclear spin directions are independant from each others \cite{Merkulov,Braun}:

\begin{eqnarray}
	\Delta_{\Omega_N}^2=\frac{2}{3} \left\langle \left\langle \bm\Omega_N^2\right\rangle\right\rangle	
\end{eqnarray}\begin{eqnarray}	
	\Delta_{\Omega_N}^2=\frac{1}{\hbar^2} \frac{n^2}{3N_L} \sum_j I^j(I^j+1) A_j^2          
\end{eqnarray}

where $N_L$ is the number of nuclei interacting with the electron in a dot and $n$ is the number of atoms in the lattice unit cell. The symbol $\left\langle\left\langle \ldots\right\rangle\right\rangle$ stand for averaging on the different nuclear field configurations. 
It is thus possible to define a characteristic dephasing time: $T_\Delta=\Delta_{\Omega_N}^{-1}$, (independently of the g-factor tensor).
The equation of motion of an average spin $\left\langle \hat{\bf S}(t)\right\rangle$ in a fixed magnetic field in a QD is then given by:

\begin{eqnarray}
\label{Smotion}
\nonumber	\left\langle \hat{\bf S}(t)\right\rangle=(\textbf S_0 .\textbf{n})\textbf{n}+[\textbf S_0-(\textbf S_0 . \textbf{n})\textbf{n}]cos(\Omega~t)\\+[\textbf S_0-(\textbf S_0 . \textbf{n})\textbf{n}]\times \textbf{n}~ sin(\Omega~t)	
\end{eqnarray}

where: $\bm\Omega=\bm\Omega_N + \bm\Omega_{ext}$,
 $\textbf{n}=\bm\Omega/\left|\bm\Omega\right|$ is a unit vector, $\bm\Omega_N$ is the nuclear precession vector in a given QD as defined in (\ref{ON}) and $\textbf{S}_0$ is the initial average electron spin \cite{Merkulov}. 
$\Omega_{ext}= g_{eff} \mu_B B_{ext}/\hbar$ represents the contribution of the external field to the total precession vector $\bm\Omega$. In this expression, an effective Land\'e g-factor has been introduced, which is defined by : $g_{eff}^2=g_\bot^2\sin^2\eta+ g_\parallel^2\cos^2\eta$, where $\eta=(\textbf e_z,\textbf B_{ext})$ is the angle of the external magnetic field with the $Oz$ axis. The angle $\eta'=(\textbf e_z,\bm\Omega_{ext})$ is given by $\eta'=Arctan(\frac{g_\bot}{g_\parallel}tan~\eta)$.

Expression (\ref{Smotion}) is then averaged over the QD ensemble, taking into account the Gaussian variations of $\bm\Omega_N$ characterized by the equation (\ref{OmegaN}) \cite{Braun, Merkulov}.
The calculation is performed in the Appendix for an arbitrary orientation of the external magnetic field (the $Ox$ axis is chosen in the plane defined by $S_0$ and $B_{ext}$). We assume that no dynamical polarization of the nuclei occurs; so that the ensemble average of $\bf\Omega_N$ is zero. For a pure transverse external field (Voigt configuration) this assumption is always valid, since no dynamical polarization of the nuclei (i.e. Overhauser effect) can occur \cite{Paget} \footnote{If $\bf B_{ext}$ has a longitudinal component, dynamical polarization of the nuclei may occur. In that case, the calculation can be performed by adding some static contribution to $\bm\Omega_{ext}$, corresponding to the ensemble average of the nuclear precession vector \cite{Paget}.}.

The average electron spin $\textbf S(t)=\left\langle 	\left\langle \hat{\bf S}(t)\right\rangle\right\rangle$ can be expressed as the sum of two contributions $ \textbf S(t)=\textbf S^\infty+ \textbf S_1 (t)$, where $S^\infty$ is time independant and $\textbf S_1 (t)$ contains the oscillating contribution which damps as $t$ increases. The expressions $\textbf S^\infty$ and $\textbf S_1 (t)$ are given respectively by the expressions (\ref{Sinf}) and (\ref{S1sans}).

For weak magnetic fields ($B_x < 200 mT$), taking $T_\Delta=500ps$ as deduced from the measurement at zero magnetic field \cite{Braun} leads to satisfactory fits of the experimental data. However, for larger magnetic fields, we will see below that the contribution of the nuclear field fluctuations is not sufficient to explain the observed damping (see Fig.\ref{fig:Compare}). Moreover, the dephasing time $T_\Delta$ is field dependent, which contradicts the fact that, under transverse magnetic field, the nuclear field fluctuations are in principle insensitive to the applied magnetic field.

\begin{figure}
\includegraphics[width=0.47\textwidth]{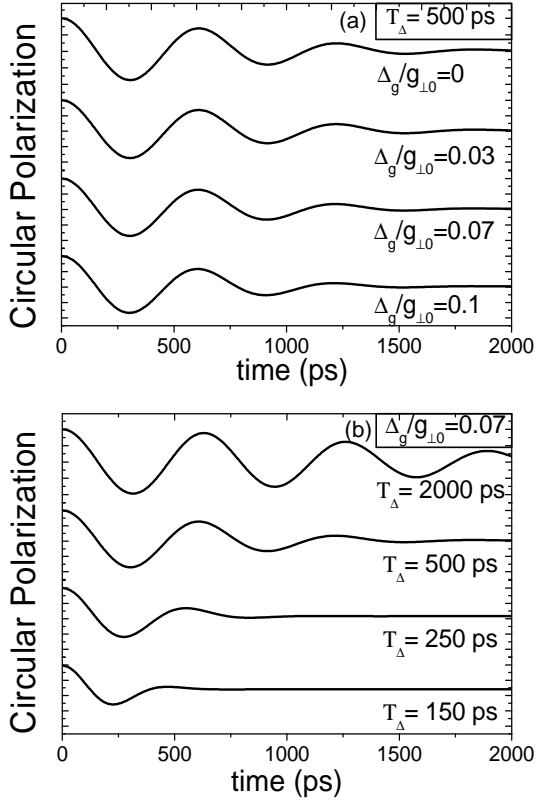}
\caption{\label{fig:theories} Theoretical curves from the model developped in the appendix showing the spin quantum beats in a weak transverse magnetic field ($B_x=150mT$)when the amplitude of the $g$-factor and the nuclear field fluctuations vary. (a) fixed nuclear field fluctuations ($T_\Delta=500ps$) and increasing $g$ fluctuations. (b) fixed $g$-factor fluctuations ($\Delta_g/g_{\bot0}=0.07$) and increasing the nuclear field fluctuations (i.e. decreasing the parameter $T_\Delta$. }
\end{figure} 

\begin{figure}
\includegraphics[width=0.47\textwidth]{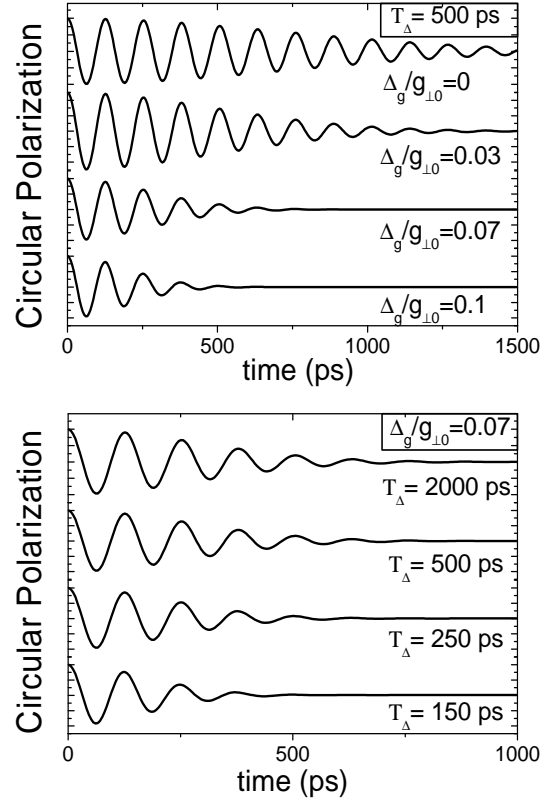}
\caption{\label{fig:theories2} Same as Fig.\ref{fig:theories} but for a stronger transverse magnetic field ($B_x=750mT$) .}
\end{figure}

In fact, the magnetic field dependent damping arises from variations of the electron g-factor over the QD ensemble, leading to a spreading of the Larmor frequencies with increasing $B_x$ \cite{Dutt, Greilich}. For instance, the origin of this inhomogeneity could come from the different dot sizes, or chemical repartition of In in the QDs. This contribution is the second source of spin dephasing for an ensemble of quantum dots. Under the reasonable assumption that the nuclear field and g-factor fluctuations are uncorrelated, we have included the variations of the latter in the model of Merkulov \textit{et al.} \cite{Merkulov}.
For an arbitrary magnetic field orientation, we did the simplifying assumption that the transverse and longitudinal fluctuations ($\delta g_\bot$ and $\delta g_\parallel$ respectively) are strongly correlated, according to 

\begin{eqnarray}
	\frac{\delta g_\bot}{\delta g_\parallel} \approx  \frac{g_\bot}{g_\parallel} \approx \frac{g_{\bot0}}{g_{\parallel0}}
\end{eqnarray}
where $g_{\bot(\parallel) 0}$ is the average $g_{\bot(\parallel)}$ value.
Under this assumption, only the amplitude of $\Omega_{ext}$ fluctuates from dot to dot, but not its direction, which is given by the angle $\eta'=\arctan(\frac{g_{\bot 0}}{g_{\parallel 0}}~tan~\eta)$. This approach yields exact results for pure transverse or pure longitudinal magnetic fields. For oblique fields, it gives some interpolation between the two cases. 
The fluctuations of the precession vector due to external magnetic field are now given by the gaussian distribution: 

\begin{eqnarray}
\label{OmegaE}
W(\Omega_{ext})=\frac{1}{\sqrt{2\pi}\Delta_{\Omega}} exp\left(-\frac{(\Omega_{ext}-\Omega_0)^2}{2\Delta_\Omega^2}\right)
\end{eqnarray}

where $\bm\Omega_{ext}=\Omega_{ext} ~\textbf e_\eta'$, $ \Omega_0 = \mu_B(g_{\bot0}^2\sin^2\eta+ g_{\parallel0}^2\cos^2\eta)^{1/2} B_{ext}$ and  $\Delta_\Omega$, defined by $\Delta_\Omega^2=\Delta g_\bot^2 sin^2\eta+ \Delta g_\parallel^2 cos^2\eta$, characterizes the fluctuation of the precession vector length. Note this corresponds to a gaussian probability distribution characterized by the parameter $\Delta_{g_{eff}}$, and the average value $g_{eff,0}$.

Then expression (\ref{Smotion})  can be averaged taking both gaussian variations for $B_N$ and $g_\bot$. The
new equations for $\textbf S^\infty$ and $\textbf S_1 (t)$ are now given respectively by the expression ($\ref{Sinf}$) taken for $g_{eff,0}$ and ($\ref{S1avec}$), from which the time resolved circular polarization can be deduced.

Let us now explain the different effects of the nuclear field fluctuations and the g-factor fluctuations. 
The obtained theoretical curves derived from the model developped in the appendix are shown in Fig.\ref{fig:theories} for $B_x = 150mT$ and in Fig.\ref{fig:theories2} for $B_x = 750mT$. 
In Fig.\ref{fig:theories}(a) and Fig.\ref{fig:theories2}(a), the nuclear field fluctuations are kept constant ($T_\Delta=500ps$), while the $g$-factor fluctuations increase. In Fig.\ref{fig:theories}(b) and Fig.\ref{fig:theories2}(b) $\Delta_g/g_{0\bot}$ is kept constant at 0.07 while the nuclear fluctuations increase. From the comparison of both figures it is clearly seen that, at weak transverse magnetic field, the damping of the oscillations is mostly determined by the nuclear field fluctuations, while both contributions are necessary under larger magnetic field using realistic sets of parameters.

Note that the curve in Fig.\ref{fig:theories}(a) for g-factor fluctuations $\Delta_g/g_{\bot0}=0$ allows us to extract the contribution of the hyperfine interaction (the equation of the spin quantum beats is given by the expressions (\ref{Sinf}) plus (\ref{S1sans})). We clearly see that the nuclear field fluctuations lead to a first contribution to damping of the $P_c$ oscillations even for the strongest $B_x$ applied. The transverse magnetic field $B_x$ is responsible for a large amplitude of spin quantum beats and yields a precise measurement of the spin decoherence time $T_2^*$. The characteristic dephasing time appearing in the model at zero magnetic field is $T_\Delta=\Delta_{\Omega_N}^{-1}$ which typical value is here $T_\Delta=500 ps$. The characteristic time for the decay of the oscillations amplitude is $2T_\Delta$ (see equation (\ref{S1sans})).

The variations of $P_c(t)$ according to the model including the g-factor fluctuations are displayed on figure \ref{fig:Compare}, for $\Delta_g/g_{\bot0}=0.07$ and $\Delta_g/g_{\bot0}=0$ for comparison. We see that, as expected, the g factor fluctuations lead to an increase of the dephasing. The new characteristic dephasing time $T_2^*$ for an ensemble of dots is now given by:

\begin{eqnarray}
\label{Dephasing}
	T_2^*=\frac{T_\Delta}{\left(1+2\left(\frac{\Delta_{g_{eff}}}{g_{eff,0}}\frac{B}{\Delta_B}\right)^2\right)^{1/2}} 
\end{eqnarray}

where $\Delta_B=\hbar/g_{eff} \mu_B T_\Delta$. The variations of $T_2^*/T_\Delta$ with the magnetic field are displayed on figure \ref{fig:theory} (bold line). When the magnetic field is much stronger than the nuclear field fluctuations ($B>>\Delta_B .g_{eff,0}/\Delta_{g_{eff}}$), these variations can be approximated by :
\begin{eqnarray}
	 T_2^* = \frac{1}{\sqrt{2}} \frac{g_{eff,0}}{\Delta_g} \frac{\Delta_B}{B} T_\Delta
\end{eqnarray}

\begin{figure}
\includegraphics[width=0.47\textwidth]{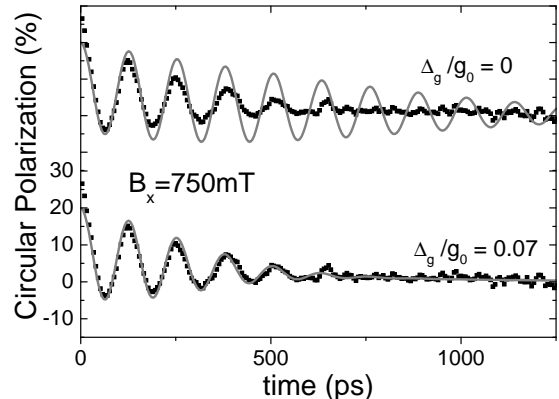}
\caption{\label{fig:Compare} Sample 1: comparison of experimental (dotted line) and theoritical curves of time resolved circular polarization for $B_x=750mT$. The theoretical curves (grey line) is given with ($\Delta_g/g_{0\bot}=0.07$) or without ($\Delta_g/g_{0\bot}=0$) g-factor fluctuations. Here $T_\Delta=500ps$ and $T_x^*=400ps$ (see text). }
\end{figure} 

The corresponding dotted line in figure \ref{fig:theory}, which is the expression taken in ref. \cite{Greilich}, is given for comparison.

At this stage, the model correctly describes the period and the damping of the electron spin quantum beats. For transverse magnetic field comparable to the nuclear field fluctuation the beats are non symetrical with respect to $P_c=0$. However, in the investigated range, $B_x$ is stronger than $\Delta_B$, which result in symetrical $P_c(t)$ oscillations. Thus, the observed asymmetry  remains to be explained. 

We believe that the origin of this asymmetry lies in the QDs charge fluctuations. Besides dots doped with a single resident hole, some are neutral and some contain two resident holes. So that neutral excitons ($X^0$) and doubly charged excitons ($X^{2+}$) are detected. 
The anisotropic exchange interaction (AEI) between electron and hole in $X^0$ and $X^{2+}$ would lead in principle to beats of $P_c(t)$ for these QDs even for $B_x=0$ \cite{Senes, Laurent}. However, the latter quickly damp due to the strong dispersion of the exchange energy from dot to dot \cite{Bayer,Patton,Flissikowski, Langbein}. 
For $X^0$ and $X^{2+}$, the spin polarization decay can be described by: $\textbf{S}(t)=S_0 e^{-t/T_x^*}$ where $T_x^*$ is an ensemble inhomogeneous dephasing time. This expression is valid provided that the dispersion of the AEI energy is of same order of magnitude as its average value. 
Defining $\alpha$ as the relative number of dots with charge state other than one, the time evolution of the circular polarization is given by: $P_c(t)=-2S_0\left((1-\alpha) S(t)/S_0 +\alpha e^{-t/T_x^*}\right)$ \footnote{We consider here that the lifetime of the radiative transition is the same for $X^0$, $X^+$ or $X^{2+}$, and that the electron-hole anisotropic exchange interaction is comparable for $X^0$ and $X^{2+}$.}.
The theoretical curves of $P_c(t)$ on Fig.\ref{fig:Compare} (grey lines) are displayed using the
parameters $T_\Delta=500ps$, $\alpha=0.4$ and $T_x^* = 400ps$, with ($\Delta_g/g_{0\bot}=0.07$) and without ($\Delta_g/g_{0\bot}=0$) g-factor fluctuations and are compared to the experimental curve (dotted line). 

\begin{figure}
\includegraphics[width=0.47\textwidth]{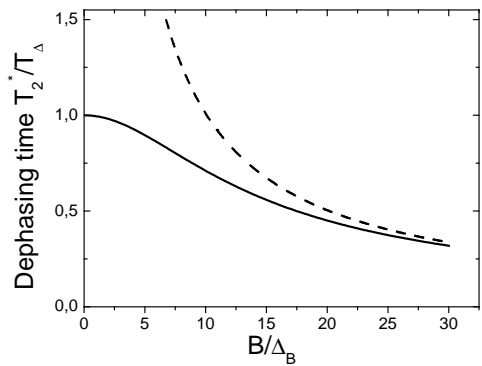}
\caption{\label{fig:theory} Dephasing time as function of the relative magnetic field. Dotted line shows the exponential model used in ref \cite{Greilich}. Bold line represent our calculations (see equation \ref{Dephasing}.) }
\end{figure} 

To summarize, the only adjustable parameters of the model are $\Delta_g/g_{\bot 0}$, $T_x^*$, and the weight $\alpha$ of QDs with 0 or 2 holes (depending on the doping distribution), all independent of the magnetic field. The nuclear field fluctuations in the QDs ensemble are determined from a fit of the circular polarization without magnetic field \cite{Braun}, and the average transverse g-factor $g_{\bot0}$ from the period of the oscillations (see fig.1). Note that $\Delta_g/g_{\bot0} = 0.07$ is consistent with previous studies \cite{Bayer2}. 

The curves on Figures \ref{fig:5269} and \ref{fig:5274} are fitted using a unique set of parameters $\Delta_g/g_{\bot 0}$, $T_x^*$, changing only the magnetic field. The parameter $\alpha$ takes into account that sample 2 ($N_A=15\times10^{10}cm^{-2}$) is doped more than sample 1 ($N_A=5\times10^{10}cm^{-2}$). We take $\alpha=0.4$ for sample 1 and 0.6 for sample 2. We observe a good agreement between theory and experiment for several applied transverse magnetic fields.


In conclusion, we have developed a model to describe electron spin dynamics in an ensemble of p-doped
QDs under transverse magnetic fields. We have shown that: (i) the hyperfine interaction,  (ii) g-factor fluctuations from dot to dot, and (iii) QD charge variations, have to be taken into account to explain the time evolution of the emited circular polarization $P_c$. The experiments presented here demonstrate the key role played by the nuclear field fluctuations
on the electron spin coherence dynamics in QDs and we found a dephasing time of about
$1~ns$ for this contribution. Note that in single QD experiments, only the contribution due to the interaction with the nuclei should be observed.

\begin{acknowledgments}
We are grateful to K. V. Kavokin for fruitful discussions and to A. E. Zhukov, V.M. Ustinov and V. K. Kalevich for sample growth and preparation. We thank MOMES ANR and FSE for financial support. X. Marie acknowledges the support from Institut Universitaire de France.
\end{acknowledgments}

\appendix

\section{Theoretical Model}
We first take into account only the fluctuations of the hyperfine field represented by
the equation (\ref{OmegaN}). 

The calculation is performed in such a $(OXYZ)$ frame that the precession vector $\bm\Omega_{ext}$ is along the $OZ$ axis ($\eta'=(\textbf e_z,\textbf e_Z)$) and the initial electron spin axis $\textbf{S}(0)$ is in the $(XOZ)$ plane, i.e. $OX$ belongs to the $(\textbf{S}(0),\bm\Omega_{ext})$ plane. 

Averaging the equation (\ref{Smotion}) of the spin motion we can find the evolution of the mean spin as a function of time, we obtain:

\begin{eqnarray}
	\textbf{S}(t)=\textbf S^{\infty} + \textbf S_1(t)
\end{eqnarray}

The first contribution is time independant. It represents the asymptotic limit for $\textbf{S}(t)$ when $t\rightarrow\infty$. The different values of $\textbf{S}(t)$ in the $(OXYZ)$ frame are:

\begin{eqnarray}
\label{appendix1}
 \nonumber S_X^\infty=\frac{S_{0X}}{2\beta^2}\left(1-\frac{1}{\sqrt\pi}\int^{\infty}_{-\infty} e^{-(z^2+\beta^2)} \frac{e^{2\beta z}-1}{2\beta z}dz\right) 
\end{eqnarray}\begin{eqnarray}
 S_Y^\infty=0
\end{eqnarray}\begin{eqnarray}\nonumber
 S_Z^\infty= \frac{S_{0Z}}{\beta^2}\left(\beta^2-1+\frac{1}{\sqrt\pi}\int^{\infty}_{-\infty} e^{-(z^2+\beta^2)} \frac{e^{2\beta z}-1}{2\beta z}dz\right)   
\end{eqnarray}

We use in this appendix $\beta=\frac{\Omega_{ext}}{\Delta_{\Omega_N}}$, $z=\frac{\Omega}{\Delta_{\Omega_N}}$, $\tau=\frac{t}{T_\Delta}$ as reduced parameters in units of the nuclear contribution to the precession vector fluctuation. The parameter  $z$ is taken algebric and we recall that $\bm\Omega=\bm\Omega_{ext}+\bm\Omega_N=\Omega.\textbf{n}$.

Note that the relation:
\begin{eqnarray}
	2 \frac{S_X^{\infty}}{S_{0X}}+\frac{S_Z^{\infty}}{S_{0Z}}=1
\end{eqnarray}
is fulfilled (see ref. \cite{Merkulov}) whatever the value of $\beta$ is. 

The second contribution contains the oscillating contributions which damp with a typical time $2T_\Delta$. We define the following integrals: 
\begin{eqnarray}
	I(\beta,\tau)=\frac{e^{\tau^2/4-\beta^2}}{\sqrt \pi} \int^{\infty}_{-\infty} e^{-z^2} \frac{e^{2\beta z}-1}{2\beta z} cos\left(z\tau\right)dz
\end{eqnarray}
\begin{eqnarray}\nonumber
	I(\beta,\tau)=\frac{1}{2\beta}( cos(\beta\tau) I_2(\beta,\tau)
	\\
	+\frac{\tau}{2}sin(\beta\tau) I_1(\beta,\tau)-e^{-\beta^2} I_2(0,\tau))
\end{eqnarray}

using the the following auxilliary integrals:
\begin{eqnarray}
	I_1(\beta,\tau)=\frac{1}{\sqrt \pi}\int^{\infty}_{-\infty}\frac{e^{-(z-\beta)^2}}{z^2+\tau^2/4}dz
\end{eqnarray}\begin{eqnarray}
	I_2(\beta,\tau)=\frac{1}{\sqrt \pi}\int^{\infty}_{-\infty}\frac{ze^{-(z-\beta)^2}}{z^2+\tau^2/4}dz
\end{eqnarray}

We can thus rewrite the different components of $\textbf S^\infty$ as:

\begin{eqnarray}
\label{appendix1}
 \nonumber  S_X^\infty=\frac{S_{0X}}{2\beta^2}\left(1-I(\beta,0)\right)
\end{eqnarray}\begin{eqnarray}
 S_Y^\infty=0
\end{eqnarray}\begin{eqnarray}\nonumber
 S_Z^\infty= \frac{S_{0Z}}{\beta^2}\left(\beta^2-1+I(\beta,0)\right)  
\end{eqnarray}

and thus:
\begin{eqnarray}
\label{sinf}
 \nonumber  S_X^\infty=\frac{S_{0X}}{2\beta^2}\left[1-\frac{1}{2\beta}\left(I_2(\beta,0)-e^{-\beta^2}I_2(0,0)\right)\right]
\end{eqnarray}\begin{eqnarray}\label{Sinf}
 S_Y^\infty=0
\end{eqnarray}\begin{eqnarray}\nonumber
S_Z^\infty= \frac{S_{0Z}}{\beta^2}\left[\beta^2-1+\frac{1}{2\beta} \left(I_2(\beta,0)-e^{-\beta^2}I_2(0,0)\right)\right]
\end{eqnarray} 

The oscillating contributions $\textbf S_1(t)$ are given by the following expressions:
\begin{widetext}
\begin{eqnarray}
\label{appendix2}\nonumber 
S_{1X}(t)=S_{0X}e^{-\tau^2/4}\left\{(1-\frac{1}{2\beta^2})cos(\beta\tau) - \frac{\tau}{2\beta} sin(\beta\tau)+\frac{1}{2\beta^2} I(\beta,\tau)\right\}
\end{eqnarray}\begin{eqnarray}\label{S1sans}
S_{1Y}(t)=-S_{0X}e^{-\tau^2/4}\left\{(1-\frac{1}{2\beta^2})sin(\beta\tau) + \frac{\tau}{2\beta} cos(\beta\tau)\right\}
\end{eqnarray}\begin{eqnarray}\nonumber
S_{1Z}(t)= \frac{S_{0Z}}{\beta^2}e^{-\tau^2/4}\left\{cos(\beta\tau)-I(\beta,\tau)\right\}
\end{eqnarray} 
\end{widetext}

Note that the following relation holds:
\begin{eqnarray}
	\frac{d}{dt}\left[\frac{2 S_{1X}(t)}{S_{0X}}+\frac{S_{1Z}(t)}{S_{0Z}}\right]=2\Omega_{ext}\frac{S_{1Y}(t)}{S_{0X}}
\end{eqnarray}
For the integral $I(\beta,0)$ the series expansion can be given:
\begin{eqnarray}
	I(\beta,0)=e^{-\beta^2}\sum_{p=0}^\infty \frac{\beta^{2p}}{p!(2p+1)}
\end{eqnarray}
From this expression we can calculate $I(0,0)$ and find the same limit as in Ref.\cite{Merkulov}:
\begin{eqnarray}
	S^\infty_X\approx\frac{1}{3}S_{0X};~~ S^\infty_Y=0; ~~S^\infty_Z\approx\frac{1}{3}S_{0Z}
\end{eqnarray}

For small magnetic field ($\beta<<1$), the general expression of $\textbf S(t)$ is, up to second order with respect to $\beta$ :
\begin{eqnarray}
\label{appendix1}\nonumber 
 S_X(t)= \frac{S_{0X}}{3}\left\{1+2e^{-\tau^2/4}(1-\frac{\tau^2}{2})\right\}+O(\beta^2)
\end{eqnarray}\begin{eqnarray}
 S_Y(t)=0
\end{eqnarray}\begin{eqnarray}\nonumber
 S_Z(t)= \frac{S_{0Z}}{3}\left\{1+2e^{-\tau^2/4}(1-\frac{\tau^2}{2})\right\}+O(\beta^2)
\end{eqnarray} 

An other limit case is found  at high magnetic field ($\beta>>1$). The expression for $\textbf S^\infty$ is:
\begin{eqnarray}
\nonumber
  S_X^\infty\approx 0; ~~S_Y^\infty\approx 0 ~~S_Z^\infty\approx S_{0Z}
\end{eqnarray}

and the expression for  $\textbf S_1(t)$ is:
\begin{eqnarray}
\label{appendix3}\nonumber 
S_{1X}(t)=S_{0X}e^{-\tau^2/4}cos(\beta\tau) 
\end{eqnarray}\begin{eqnarray}
S_{1Y}(t)=-S_{0Y}e^{-\tau^2/4}sin(\beta\tau)
\end{eqnarray}\begin{eqnarray}\nonumber
S_{1Z}(t)= 0
\end{eqnarray} 

and finally the expression for $\textbf S(t)$:
\begin{eqnarray}
\label{appendix1}\nonumber 
S_{X}(t)=S_{0X}e^{-\tau^2/4}cos(\beta\tau) 
\end{eqnarray}\begin{eqnarray}
S_{Y}(t)=-S_{0Y}e^{-\tau^2/4}sin(\beta\tau)
\end{eqnarray}\begin{eqnarray}\nonumber
S_{Z}(t)= S_{0Z}
\end{eqnarray} 
Note that, at long time delay, we have $\textbf S_1(t)\rightarrow 0$ and  $\textbf S(t)\rightarrow S^\infty$ when $\tau\rightarrow\infty$.
\newline

Let now discuss the averaging of equations (\ref{Sinf}) and (\ref{S1sans}) if we take into account the g-factor fluctuations. 

We introduce the effective g-factor by the expression:
\begin{eqnarray}
	\frac{g_{eff}\mu_B B_{ext}}{\hbar}=\Omega_{ext}
\end{eqnarray}
where:
\begin{eqnarray}
	g_{eff}^2=g_\bot^2 sin^2\eta + g_\parallel^2 cos^2\eta
\end{eqnarray}
and $\eta=(\textbf{e}_z, \textbf{B}_{ext})$.

The g-factor fluctuations lead to $\Omega_{ext}$ variations given by the expression (\ref{OmegaE}) where:
\begin{eqnarray}
	\Delta_\Omega=\frac{\Delta_{g_{eff}}\mu_B B_{ext}}{\hbar}
\end{eqnarray}
Using the expression $\beta=\Omega_{ext}/\Delta_{\Omega_N}$ we obtain:
\begin{eqnarray}
	\Delta_\beta=\frac{\Delta_{g_{eff}}\mu_B B_{ext}}{\hbar \Delta_{\Omega_N}}
\end{eqnarray}
We assume in our model that:
\begin{eqnarray}
	\frac{\Delta_\beta}{\beta}=\frac{\Delta_{g_{eff}}}{g_{eff}}<<1
\end{eqnarray}
Moreover, if $\beta \geq\pi$ (\textit{i.e.} $B_{ext}>> h/2 g_{eff} \mu_B T_\Delta$), the envelope of the average spin oscillations decays much slower than the period of these oscillations. It follows that, in order to average $\bf{S}(t)$ ((\ref{Sinf}) and (\ref{S1sans})) over the g-factor fluctuations, we can use the following approximation: $S^\infty \approx S^\infty)_{\beta=\beta_0}$. For $S_1(t)$ we obtain the approximate expression: 

\begin{widetext}
\begin{eqnarray}
\label{appendix4}\nonumber 
S_{1X}(t)=S_{0X}e^{-\frac{\tau^2}{4}(1+2\Delta_\beta^2)}\left\{\left[1-\frac{1}{2\beta^2_0}\left(1-\frac{I_2(\beta_0,\tau)}{2\beta_0}\right)\right] cos(\beta_0\tau) - \frac{\tau}{2\beta_0} \left(1-\frac{I_1(\beta_0,\tau)}{4\beta_0^2}\right)sin(\beta_0\tau)\right\}-S_{0X}e^{-\frac{\tau^2}{4}}\frac{e^{-\beta_0^2}I_2(0,\tau)}{4\beta_0^3}
\end{eqnarray}\begin{eqnarray}\label{S1avec}
S_{1Y}(t)=-S_{0X}e^{-\frac{\tau^2}{4}(1+2\Delta_\beta^2)}\left\{\left(1-\frac{1}{2\beta_0^2}\right)sin(\beta_0\tau) + \frac{\tau}{2\beta_0} cos(\beta_0\tau)\right\}
\end{eqnarray}\begin{eqnarray}\nonumber
S_{1Z}(t)=\frac{S_{0Z}}{\beta_0^2}e^{-\frac{\tau^2}{4}(1+2\Delta_\beta^2)}\left\{\left(1-\frac{I_2(\beta_0,\tau)}{2\beta_0}\right)cos(\beta_0\tau) - \frac{\tau}{2\beta_0} \frac{I_1(\beta_0,\tau)}{2}sin(\beta_0\tau)\right\}+S_{0Z}e^{-\frac{\tau^2}{4}}\frac{e^{-\beta_0^2}I_2(0,\tau)}{2\beta_0^3}
\end{eqnarray} 
\end{widetext}

The calculation frame $(OXYZ)$ is obtained from the initial one $(Oxyz)$ by performing two rotations. One with $Oy$ axis with an angle $\eta'$ : $R_{Oy}(\eta')(Oxyz)=(Ox'y'z')$ such that $Oz'//  \bm\Omega_{ext}$. 

The second with $Oz'$ axis and angle $\phi$: $R_{Oy}(\phi)(Ox'y'z')=(OXYZ)$ such that the plane $(XOZ)$ contains $\bf S_0$. We can thus obtain the general expression of the electron spin evolution in the frame $(Oxyz)$ by:
\begin{eqnarray}\nonumber
	\begin{pmatrix} 
S_x(t)\\ 
S_y(t)\\ 
S_z(t) 
\end{pmatrix}
=
\begin{pmatrix} 
cos\eta'~ cos\phi & -cos\eta' ~sin\phi & sin\eta'\\ 
sin\phi & cos\phi & 0\\ 
-sin\eta' ~cos\phi & sin\eta' ~sin\phi & cos\eta' 
\end{pmatrix} 
\begin{pmatrix} 
S_X(t)\\ 
S_Y(t)\\ 
S_Z(t) 
\end{pmatrix} 
\end{eqnarray}

For a pure transverse magnetic field $\eta=\eta'=\pi/2$ and $\phi=0$ so the transformation takes the simple form: 
\begin{eqnarray}
\begin{pmatrix} 
S_x(t)\\ 
S_y(t)\\ 
S_z(t) 
\end{pmatrix} =
\begin{pmatrix} 
S_Z(t)\\ 
S_Y(t)\\ 
-S_X(t) 
\end{pmatrix} 
\end{eqnarray}

In addition, as here $\textbf S_0=S_{0z}~\textbf e_z$, we have: $S_x(t)=S_y(t)=0$ and $S_z(t)=-S_X(t)$. Some example are given in figure 1, figure 4 and figure 5.

\newpage 

\end{document}